# Single- and Mixed-Phase TiO$_2$ Powders Prepared by Excess-Hydrolysis of a Titanium Alkoxide

Dorian A.H. Hanaor [a]*, Ilkay Chironi[b], Inna Karatchevtseva[b],  Gerry Triani [b], and Charles C. Sorrell [a]

*[a] University of New South Wales, School of Materials Science and Engineering, Sydney, NSW  2052, Australia*

*[b] Australian Nuclear Science and Technology Organisation, Institute of Materials and Engineering Sciences, Menai, NSW  2234, Australia*

*\*Corresponding Author:  Email:  dorian@ unsw.edu.au, Ph. (+61)-404-188810*

## Abstract

To investigate excess-hydrolysis of titanium alkoxides, TiO$_2$ powders were fabricated from titanium-tetra-isopropoxide using 6:1 and 100:1 H$_2$O:Ti (r) ratios.  Powders were dried and fired at a range of temperatures ($\leq$800ºC).  Hydroxylation and organic content in powders were characterised using ATR-FTIR, laser Raman microspectroscopy, and elemental microanalysis; surface area and pore size distribution were evaluated using N$_2$ gas adsorption; phase composition was analysed using XRD and laser Raman microspectroscopy; and crystallite size was evaluated by XRD, TEM and SEM.  Results showed near-complete hydrolysis in a predominantly aqueous medium (r = 100), resulting in precipitated crystalline powders exhibiting brookite and anatase, which begin to transform to rutile below 500°C.  Powders precipitated in a predominantly organic medium (r = 6) underwent partial hydrolysis, were highly porous and exhibited an amorphous structure, with crystallisation of anatase occurring  at ~300°C and transformation to rutile beginning at 500°-600°C.

# 1. Introduction

Titanium dioxide is attracting increasing interest owing to its ability to function as a semiconductor photocatalyst. In photocatalysis, irradiation exceeding the semiconductor band gap is absorbed to generate electron-hole pairs, which can facilitate reactions of significant environmental importance, including water purification [1-4] and hydrogen production through water splitting [5-7].

The use of $TiO_2$ photocatalysts as dispersed powders provides the potential for a catalyst of higher available surface area in comparison with supported materials in the form of films or coatings. As photocatalysed reactions take place at close proximity to catalyst surfaces, $TiO_2$ powders have been used in aqueous suspensions for water purification applications [8-10] and hydrogen production [7, 11].

Titanium dioxide powders of high surface area are produced frequently by the hydrolysis of Ti alkoxides, as this approach gives rise to higher surface areas in comparison with industrially produced pigmentary $TiO_2$ [12, 13].

If no gelation-inducing reagents (hydrolysis catalysts) are used, the initial hydrolysis of titanium alkoxides results in the formation of titanium hydroxide monomers according to equation 1 [14-18], where R represents an organic chain of the formula $(C_xH_{2x+1})$:

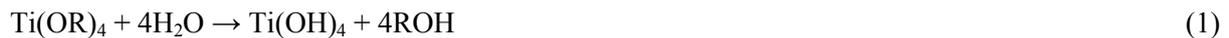

$$Ti(OR)_4 + 4H_2O \rightarrow Ti(OH)_4 + 4ROH \qquad (1)$$

Titanium dioxide with OH-terminated surfaces forms and precipitates through the condensation of hydroxide monomers to form a network of Ti-O bonds. This reaction, described in equation 2 [15-19], takes place through a nucleation and growth process:

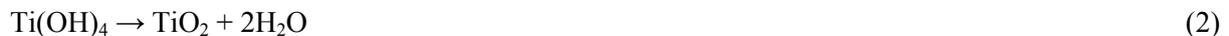

$$Ti(OH)_4 \rightarrow TiO_2 + 2H_2O \qquad (2)$$

Equation 1 generally does not proceed stoichiometrically and, in the absence of peptising agents, the rapid formation of partially hydrolysed species occurs [17, 18, 20-22]:

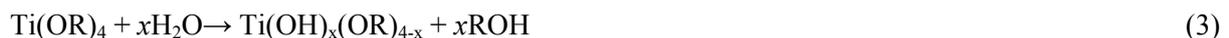

$$Ti(OR)_4 + xH_2O \rightarrow Ti(OH)_x(OR)_{4-x} + xROH \qquad (3)$$

This is accompanied by the condensation reaction and resultant precipitation described by equation 4. [23, 24]:

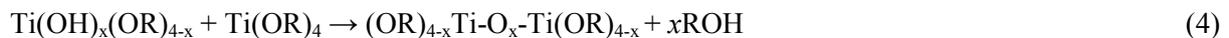

$$Ti(OH)_x(OR)_{4-x} + Ti(OR)_4 \rightarrow (OR)_{4-x}Ti-O_x-Ti(OR)_{4-x} + xROH \qquad (4)$$

In the fabrication of titanium dioxide powders, the $H_2O$:Ti ratio, known as the hydrolysis ratio *r*, is a key factor in governing the size, morphology, and crystallininity of the precipitates that form. Using low hydrolysis ratios tends to result in partial hydrolysis, with retained organic groups present in the precipitate. Higher hydrolysis ratios are necessary for complete or near-complete hydrolysis to take place [17].

The precipitation of amorphous material from solution occurs when the incipiently formed nuclei are smaller than one unit cell, while larger nuclei give rise to growth of crystalline precipitates [25]. As a larger degree of supersaturation leads to the formation of larger nuclei, excess hydrolysis above a certain ratio, reported between 6 and 200, is necessary for the precipitation of crystalline $TiO_2$ from titanium alkoxide solutions, while the use of lower hydrolysis ratios generally results in the precipitation of amorphous materials, which transform to the anatase phase through heat treatment [17, 24, 26-28].

pH levels and temperatures are critical in determining the crystalline phases of $TiO_2$ that result from precipitation [29, 30]. As with the surface area, the phase content of $TiO_2$ has a significant effect on the resultant photocatalytic performance of the material, with anatase and brookite reported to exhibit superior activity to that of the equilibrium rutile phase [31-33]. Furthermore, it is reported frequently that a mixed anatase/rutile or anatase/brookite/rutile phase composition is advantageous for photocatalytic applications owing to improved charge carrier separation [34-36].

The present work investigates the use of excess hydrolysis at near-neutral pH and its effects on the resultant surface area, organic group retention, phase composition, and morphological properties in $TiO_2$ powders formed by precipitation from a titanium alkoxide.

## 2. Experimental Procedure

### 2.1. Powder synthesis

Powders were synthesised from solutions of titanium tetra-isopropoxide (TTIP, 97%, Sigma Aldrich, USA) precipitated by hydrolysis at room temperature using predominantly organic media, and predominantly aqueous precipitation media. These two precipitation conditions were chosen as they represent the divergent approaches taken in alkoxide precipitation media employed in various studies reported in the literature [23, 27, 28, 37], resulting in conditions of moderate or high levels of excess water.

For the achievement of moderate excess hydrolysis in the present study, a 150 mL solution of 0.5 M TTIP in isopropanol was hydrolysed by the dropwise addition of a solution of distilled water in 20 mL isopropanol under rapid magnetic stirring giving a 6:1 hydrolysis ratio. This resulted in visible precipitation occurring after 1-2 min of continued stirring at a pH level of 5.6. For precipitation in a predominantly aqueous medium (with a high excess of water), a 50 mL of a 50 vol% solution of TTIP was added dropwise to 150 mL of distilled water with stirring, yielding a $H_2O:Ti$ ratio of 100:1, and resulting in rapid precipitation of dispersed agglomerates with a suspension pH level of ~5.4. Powders were denoted *r = 6*, representing those synthesised in a predominantly organic medium with 6:1 hydrolysis ratio, while *r = 100* denoted powders precipitated in a predominantly aqueous medium.

### 2.2. Heat treatment

Precipitated powders were dried at 110°C for 48 h to remove physically adsorbed water and residual isopropanol. The dried powders were placed in dense alumina crucibles and fired in air in a muffle furnace at temperatures in the range 200º-800°C for 4 h, with heating and initial cooling rates of 2°C/min.

### 2.3. Compositional analysis

The presence of hydroxyl groups and organic species in the powders was assessed using attenuated total reflection Fourier transform infrared spectroscopy (ATR FTIR) and elemental microanalysis. These analyses were done using a Spotlight 400 FTIR and Carlo Erba 1106 Microanalyser, respectively.

### 2.4. Morphological characterisation

The morphology of powders was assessed by scanning electron microscopy (SEM) using an FEI NanoSEM-230 microscope and by transmission electron microscopy (TEM) using an FEI Tecnai G2 microscope. The grain size was assessed using high-magnification SEM images (≥40,000X) using the linear intercept method while crystallite sizes were similarly assessed from TEM images.

X-ray diffraction (XRD) patterns were collected using a Philips MPD unit with CuKα radiation. XRD spectra were used to calculate the lower crystallite size limit using the Scherrer equation [38, 39]. The phase contents of the powders were determined using a slightly modified version of the method of Spurr and Myers, as shown in equation 5 [40].

$$X_{A+B} = 1 - X_R = (1 + 1.26 \frac{I_{27.5}}{I_{25.3}})^{-1} \tag{5}$$

Where the subscripts A, B, and R represent anatase, brookite, and rutile, respectively.

Whereas the original version of this equation evaluates only $X_A/X_R$ ratios, the present modified version incorporates the $X_{A+B}/X_R$ ratio, where the former is the combined anatase + brookite fraction. $I_{27.5}$ and $I_{25.3}$ are the relative intensities of the rutile (110) peak at ~27.5° 2θ and the anatase (101) at ~25.3° 2θ. Since latter is a single broad and symmetric peak, with no shoulder, it coincides with the brookite (111) peak at ~25.3° 2θ [31, 41].

Surface areas were assessed by $N_2$ adsorption at 77 K using a Micrometrics ASAP 2020 Physisorption Analyser (Brunauer Emmett Teller, BET method). These data were correlated with those determined using the Langmuir calculation method. The average pore diameters were assessed by $N_2$ adsorption at 77 K using the same instrument by the Barrett Joyner Halenda (BJH) analysis method.

## 3. Results

### 3.1. SEM micrographs

Representative SEM images, shown in **Fig. 1** and **Fig. 2**, show isotropic grains of increasing size with firing temperature. The increase in grain size and density is particularly significant between 600ºC and 800°C. Agglomeration was evident in all samples, with larger, more consolidated agglomerates, present in the r = 100 powders. A significant amount of porosity was prevalent in all samples, with larger pores visible in the well-consolidated agglomerates in the r = 100 powders fired at 700°-800°C. Samples precipitated at r = 6 exhibited less agglomeration, with a loosely packed porous structure retained after firing at 800°C, as evident from comparison of Fig. 2c with Fig. 1c

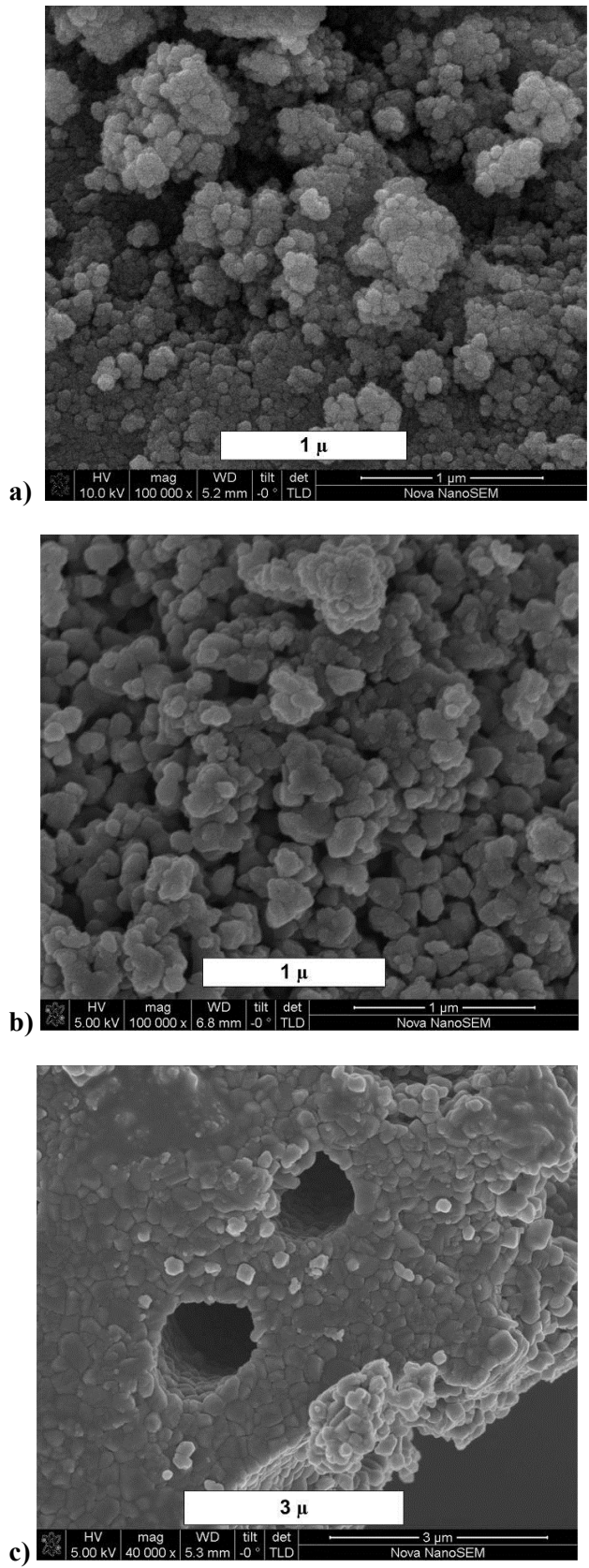

**Fig. 1.  Powders from precipitation at r = 100 fired at (a) 200°C, (b) 600°C, and (c) 800°C**

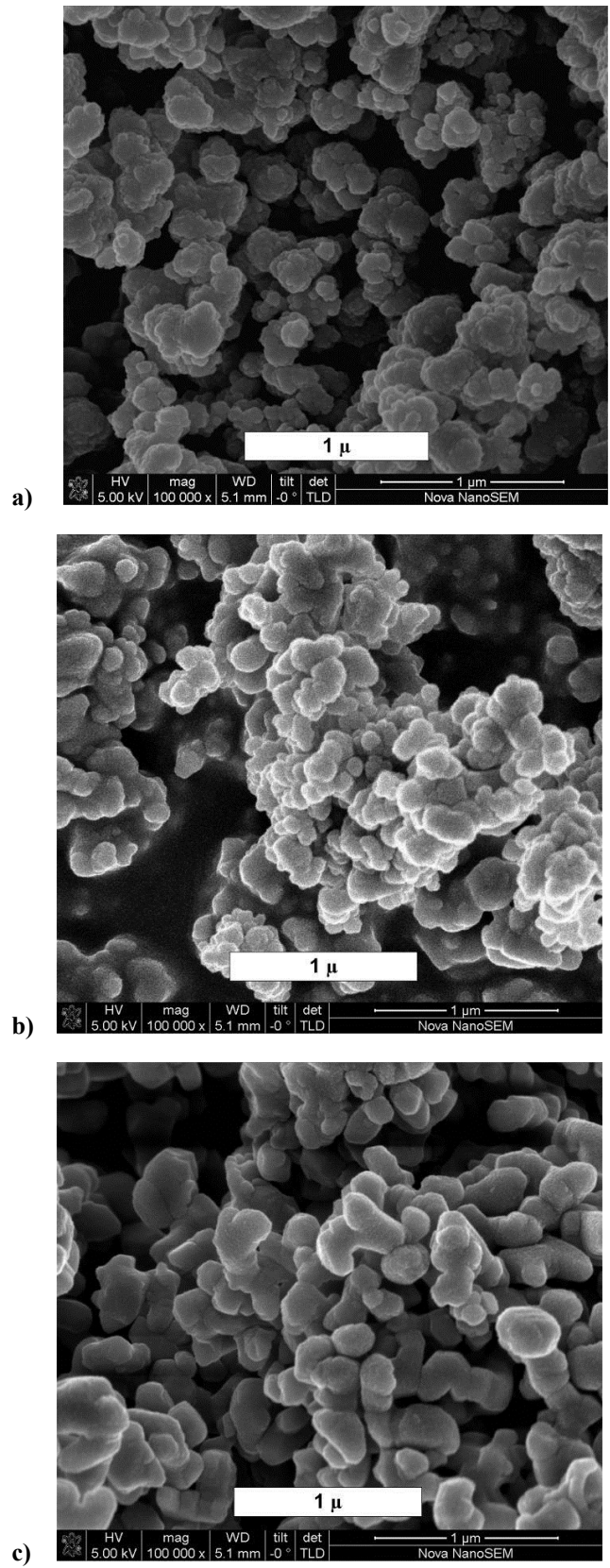

**Fig. 2. Powders from precipitation at r = 6 fired at (a) 200°C, (b) 600°C, and (c) 800°C**

### 3.2. TEM analysis

**Fig. 3 and Fig. 4** show TEM images. TEM analysis revealed that the grains visible in SEM micrographs from low temperature fired samples (≤600 °C) are comprised of finer crystallites. These crystallites exhibit significant growth in the transformation to rutile. After firing at 600°C crystallite size shows significant divergence with smaller crystallites, possibly residual anatase, coexisting alongside larger crystallites of rutile. TEM imagery is inappropriate for agglomerate size analysis as a result of sample preparation methods and the electron-opacity of larger particles.

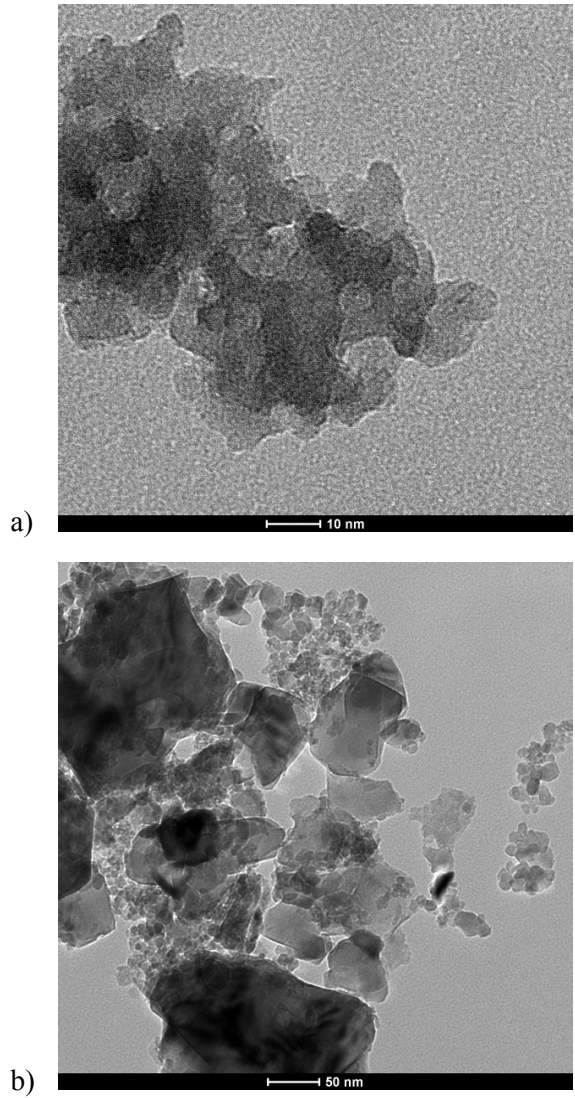

a)

b)

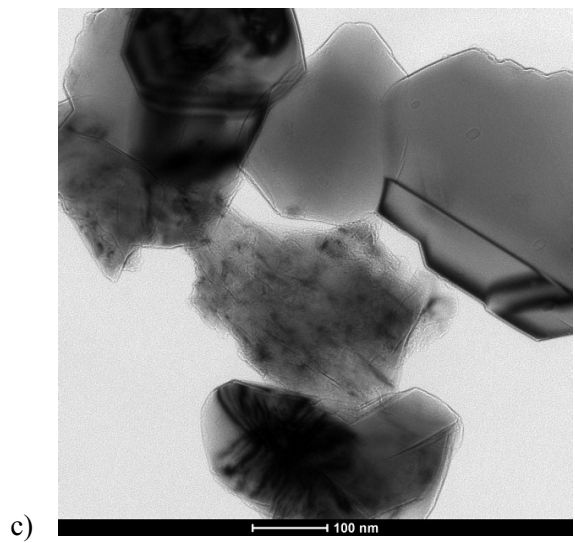

c)

**Fig. 3. TEM micrographs of r = 100 powders fired at (a) 200°C, (b) 600°C and (c) 800°C**

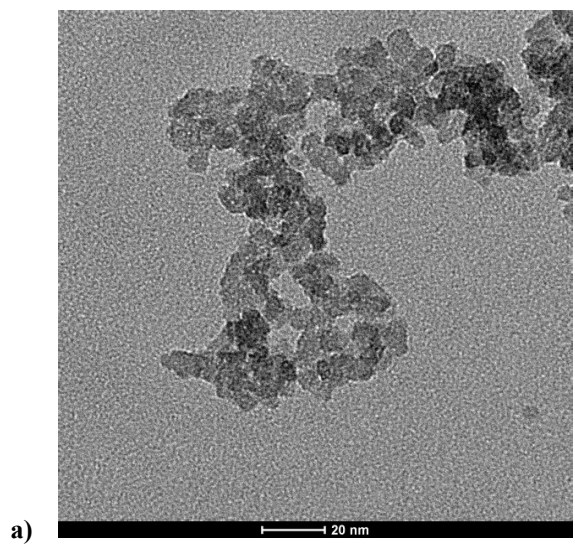

a)

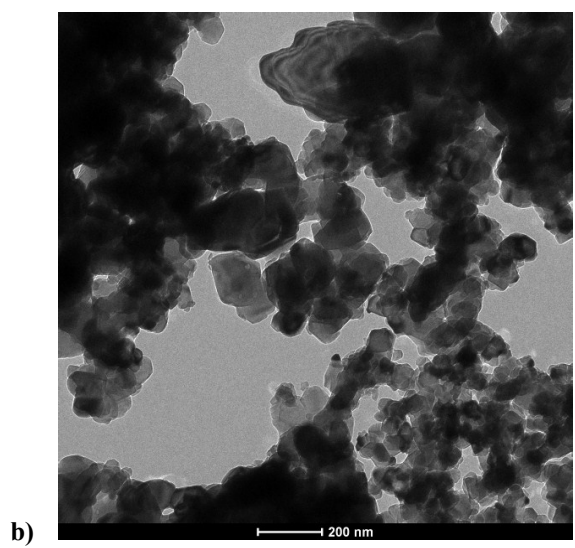

b)

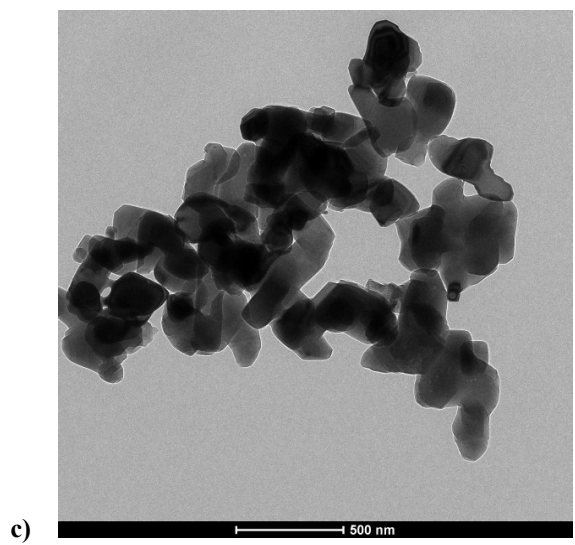

c)

**Fig. 4. TEM images of r = 6 powders fired at (a) 200°C, (b) 600°C and (c) 800°C**

### 3.3. Elemental analysis

The results for elemental analysis ,shown in **Fig. 5**, show the reduction in residual carbon content in powders with increasing firing temperature.  In comparison with powders precipitated in a predominantly aqueous medium, powders precipitated in a predominantly organic medium show higher levels of carbon owing to the presence of retained organic groups in the hydrolysis product. After firing at temperatures ≥400°C, the carbon contents approach the same baseline value.

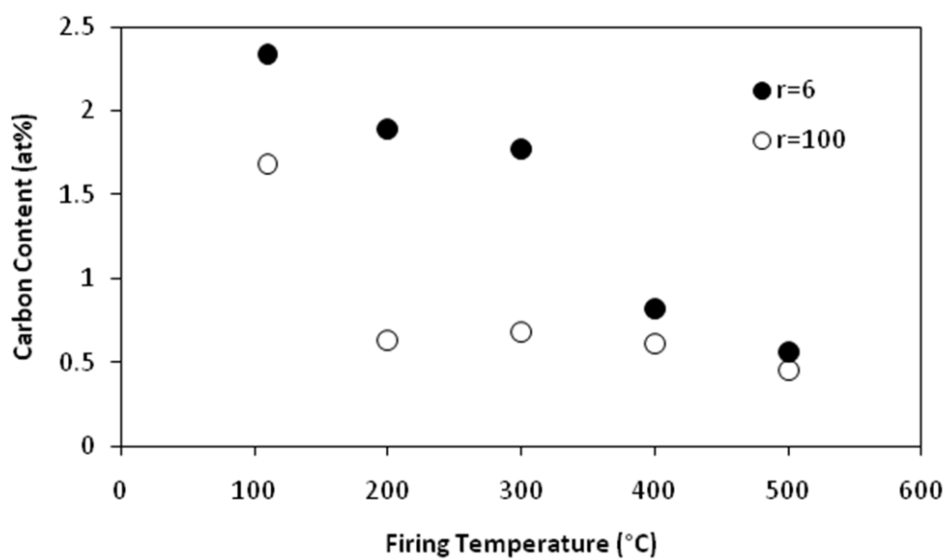

**Fig. 5.  Residual carbon content in TiO$_2$ powders as a function of firing temperature**

### 3.4. Spectroscopy

ATR-FTIR spectra, shown in **Fig. 6**, show the presence of a broad band at 3000-3600 cm$^{-1}$ corresponding to the stretching vibration of terminating hydroxyl groups in samples fired up to 500°C. The peak at 1630 cm$^{-1}$ corresponds to the bending vibration of residual surface-adsorbed H$_2$O, with the decrease in intensity with increasing temperature corresponding to the increase in bulk density (and hence decrease in surface area). Powders precipitated at r = 6 show the presence of residual organic groups after low-temperature firing, evident from the C-H stretching at 2976 cm$^{-1}$ and 900-1300 cm$^{-1}$ [23, 42].

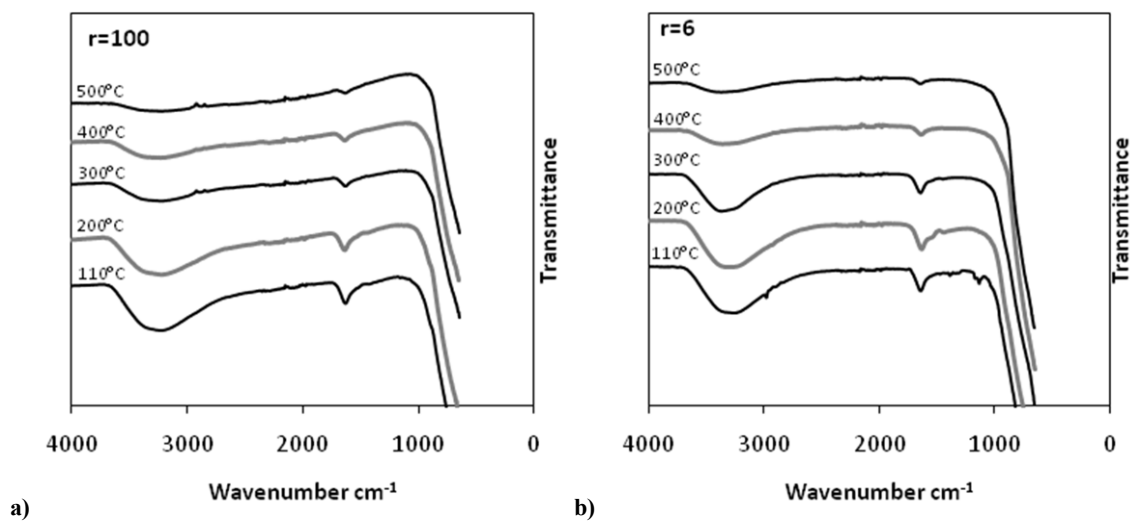

**Fig. 6. ATR-FTIR spectra at (a) r = 100 powders and (b) r = 6 powders fired at different temperatures**

Laser Raman microspectra for the low-temperature-fired samples are shown in **Fig. 7**. Similar to the ATR-FTIR spectra, peaks at 2900-3000 cm$^{-1}$ are present in r = 6 powders heated at 110°C (dried powder) and 200°C, resulting from the stretching of CH$_2$ and CH$_3$ groups. These data confirm partial hydrolysis of the alkoxide at r = 6 with the retention of organic groups in the precipitate. The broad peaks at 410 and 630 cm$^{-1}$ result from TiO$_6$ octahedral units. The high peak at 144 cm$^{-1}$ is characteristic of the anatase phase. However, short-range anatase-like order in amorphous TiO$_2$ may give rise to this peak owing to similar Ti-O bond lengths [43].

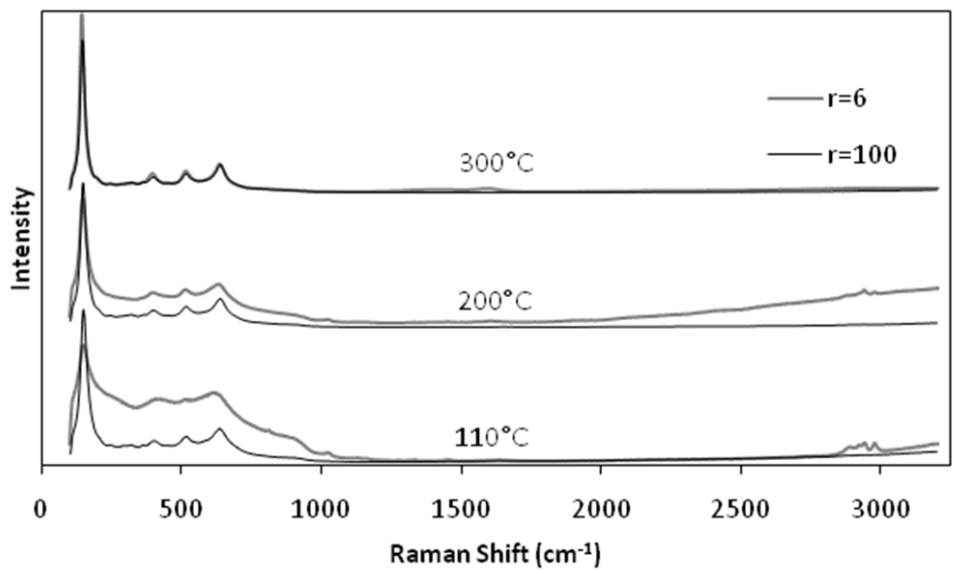

**Fig. 7. Raman spectra of powders heated at different temperatures**

### 3.5. X-ray diffraction

XRD spectra from powders precipitated at r = 100, shown in **Fig. 8**, show broad diffuse anatase peaks after firing at low temperatures owing to the presence of crystalline phases from precipitation. At samples fired at ≤500°C, the brookite (121) peak at ~30.5° 2θ is visible. The brookite (210) and (111) peaks at ~25.3 and ~25.7° 2θ coincide with the anatase (101) peak, giving rise to further peak heightening as well as broadening. As shown in **Fig. 9**, powders precipitated at r = 6 exhibit only amorphous content below 300°C, the temperature at which the anatase phase becomes observable; there is no evidence of brookite at any temperature. A small rutile (110) peak at ~27.3° 2θ is apparent at 500°C in r = 100 powders; this peak is not present at this temperature in r = 6 powders.

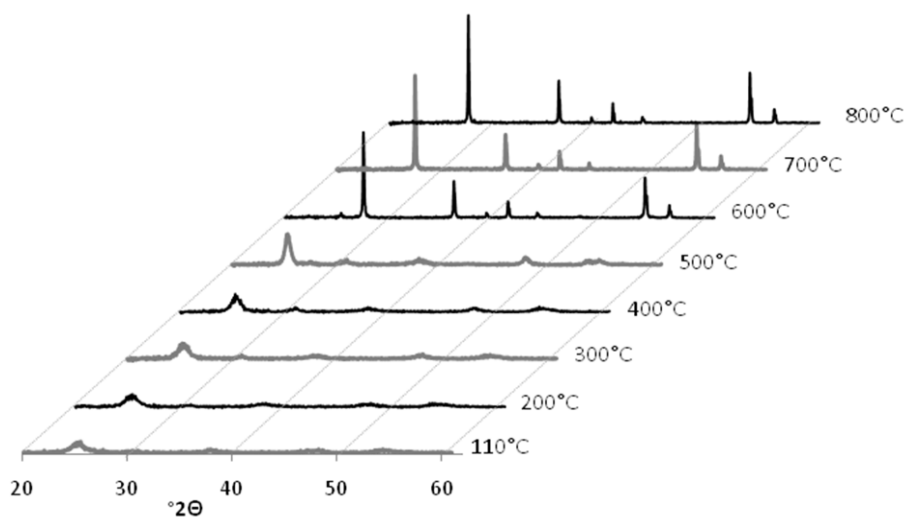

**Fig. 8.  XRD patterns of r = 100 powders fired at different temperatures**

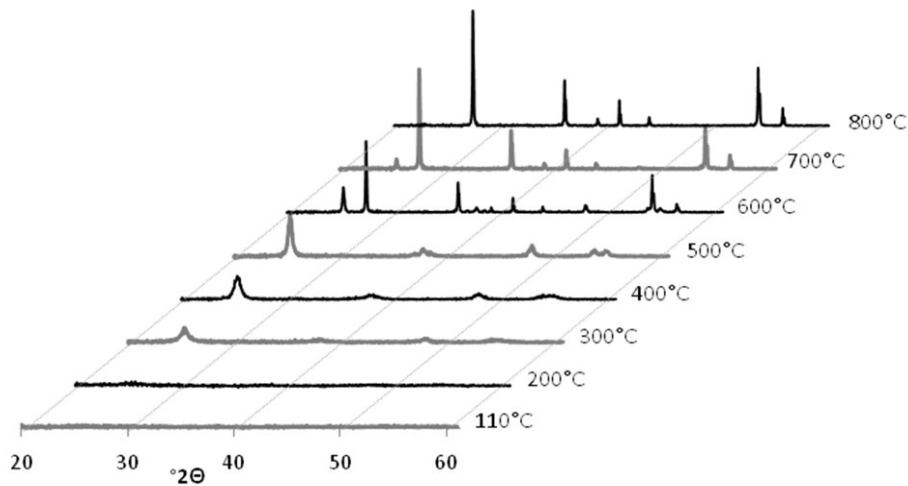

**Fig. 9.  XRD patterns of r = 6 powders fired at different temperatures**

### 3.6. Analysis of morphology and phase composition

TEM micrographs and XRD spectra were analysed to ascertain the variation in crystallite sizes with firing temperature, while the analysis of SEM micrographs shows the variation in grain size. A summary of these data is shown in **Fig. 10**.  A significant increase in both grain and crystallite size accompanies the anatase to rutile transformation at ~600°C.  As revealed by TEM analysis, grains visible in SEM analysis consist of nano-crystallites after firing at lower temperatures (≤ 500 °C). After firing at higher temperatures the grain size observed by SEM is approximately consistent with the crystallite size observed by TEM. The interpretation of XRD data using the Scherrer equation [38, 39] for morphological analysis is applicable only for crystalline materials with crystallites smaller than ~100 nm and provides an estimate of the lower limit for the crystallite size [44]. Crystallite sizes estimated by the Scherrer equation are consistent with TEM results at lower temperatures (≤ 500 °C), while at higher temperatures XRD analyses yield smaller crystallite size values. This is a result of the divergence in crystallite sizes shown also in Fig 3b and Fig. 4b along with the aforementioned ineffectiveness of XRD methods for the morphological analysis of coarse-grained material. It should be noted that after treatment at 110° and 200°C, r = 6 powders are amorphous and thus cannot be analysed for crystallite size by XRD methods. Grains of these amorphous powders are still comprised of nanoscale subgrains as evident in Fig. 2a and Fig. 4a. Grains observed by SEM as well as crystallites observed by TEM were larger in r = 100 powders fired under 500 °C in comparison with r=6 powders.

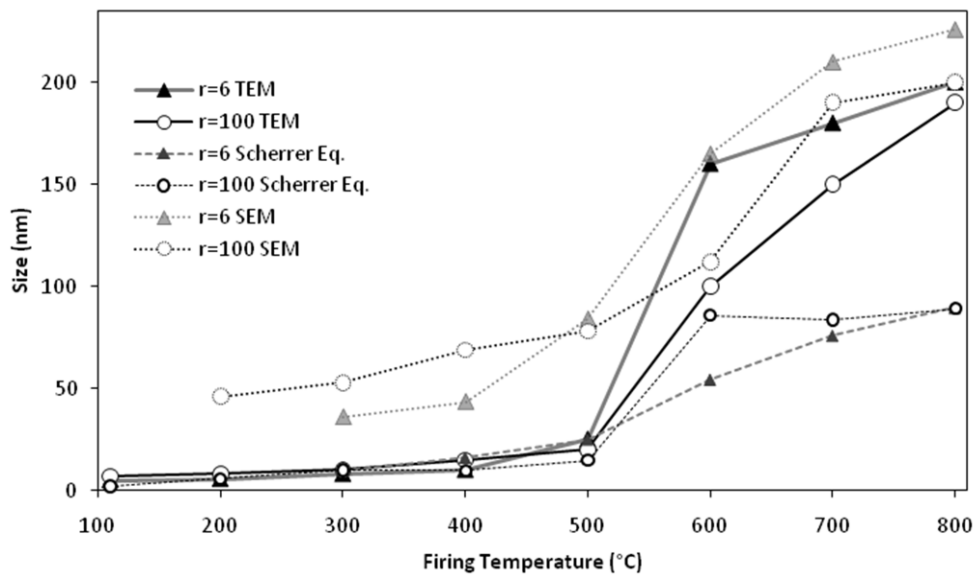

**Fig. 10. Crystallite and grain size analyses by TEM, XRD and SEM**

As shown in **Fig. 11**, the phase composition, determined from the XRD patterns using the modified method of Spurr and Myers [40] shows a more rapid transformation to rutile in samples formed from aqueous precipitation. The initial appearance of rutile appears to be at ~500°C in r = 100 powders and at ~600°C in r = 6 powders. A more accurate approximation of the phase transformation onset temperature would require extended firing durations at narrower temperature intervals [45].

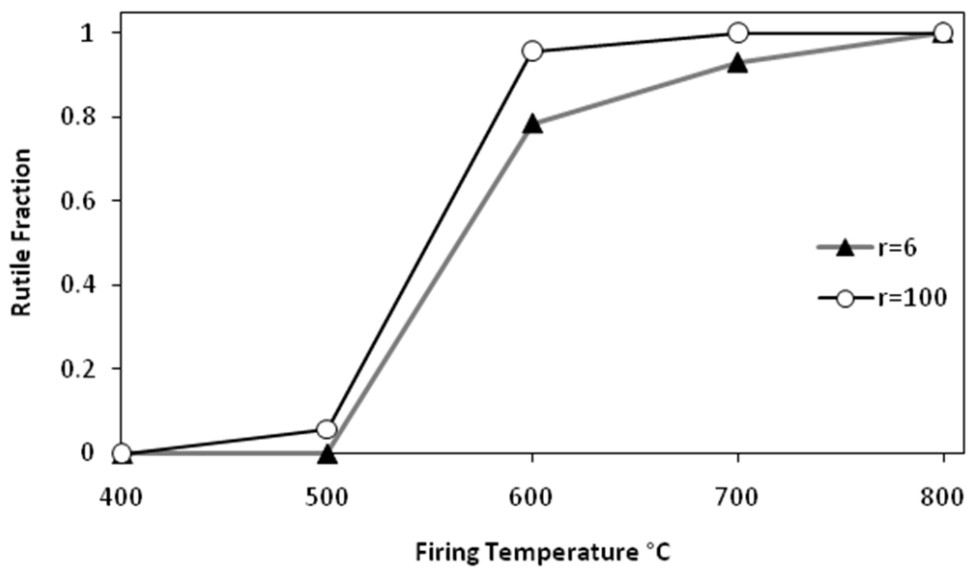

**Fig. 11. Rutile fraction calculated from XRD peaks using the method of Spurr and Myers [40]**

### 3.7. Surface area analysis

Plots of $N_2$ gas adsorption at 77 K used to determine the surface areas according to the BET method [46, 47] are shown in **Fig. 12**. In these figures, Q is the quantity of gas adsorbed in mmol·$g^{-1}$ and $P/P_0$ is the ratio of adsorptive pressure to saturation pressure. The slope and intercept of these plots were used to determine the surface area of powders, where a high value of intercept and slope translate to a low surface area. A large shift in slope and intercept of these plots is evident at a firing temperature of 600°C in r = 100 powders. This shift is present to a significantly lesser extent in powders precipitated at r = 6.

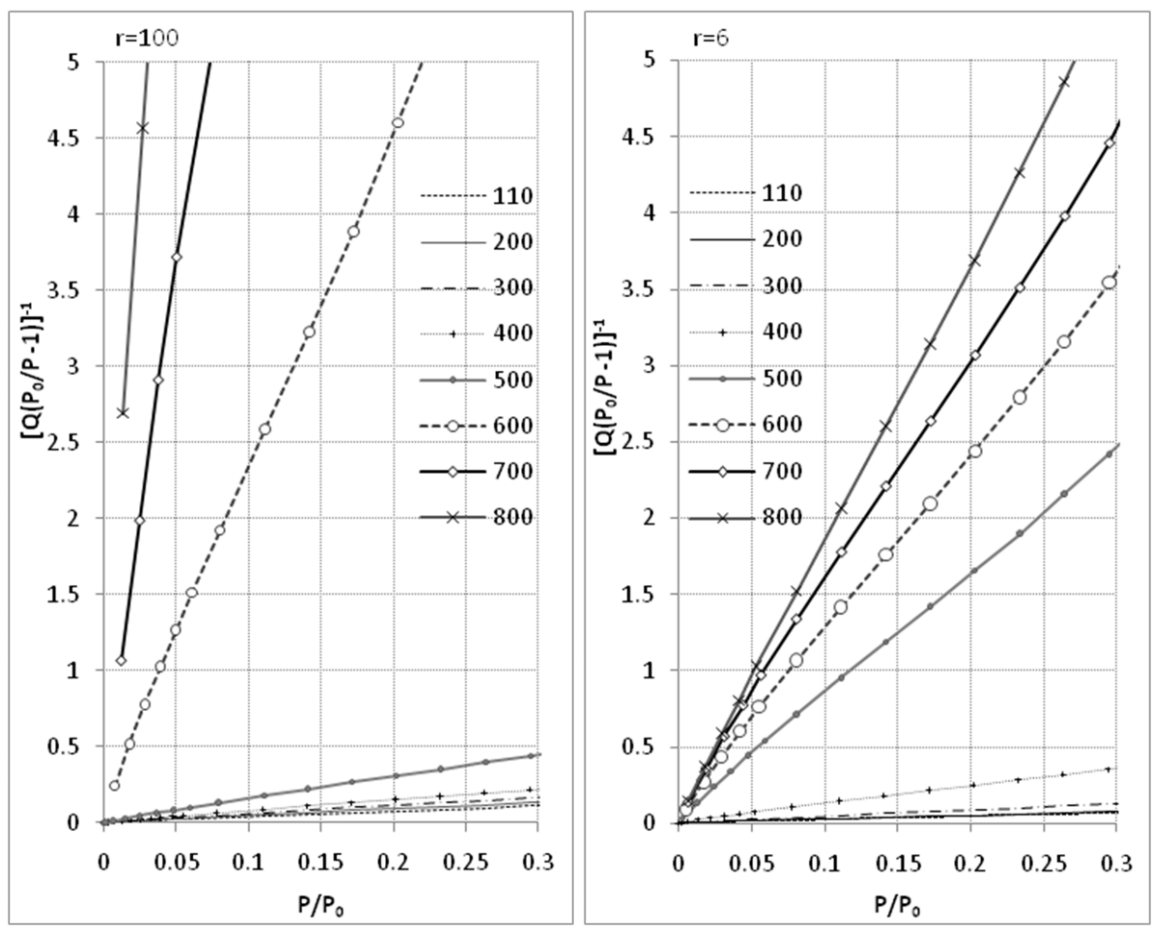

**Fig. 12.** $N_2$ adsorption at 77 K for (a) r = 100 powders and (b) r = 6 powders

The surface areas determined by these data are shown in **Fig. 13**. This figure shows an initial higher surface area in r = 6 powder in comparison with r = 100 powders, with a sharp decrease in surface area occurring with increasing firing temperature. As evident from the slopes of the $N_2$ adsorption curves in Fig. 12, the drop in surface area at 600°C is more significant in r = 100. The expanded view of the surface area after firing at 600º-800ºC shows the surface area of the r = 100 powders following the anatase to rutile transformation is significantly lower in comparison with powders precipitated in an organic medium.

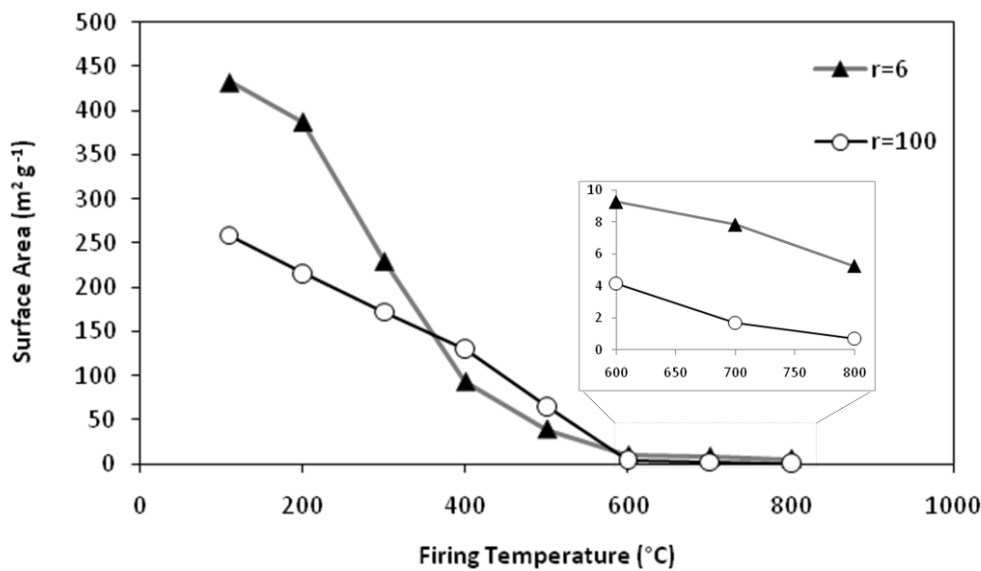

**Fig. 13. Surface areas determined by BET analysis**

The average pore diameter, determined from BJH analysis of the adsorption isotherms, exhibits an increase with firing temperature, as shown in **Fig. 14**. This is more evident in powders from aqueous precipitation, where the average pore diameter continued to increase at higher temperatures. In contrast, powders precipitated at r = 6 did not exhibit as significant an increase in pore size at high temperatures. These results are consistent with the densification behaviour of the agglomerates, which results in expansion of the inter-agglomerate pore size [48]. Hence, this effect is enhanced in the more agglomerated r = 100 powders, as indicated in SEM images shown in Figs. 1 and 2.

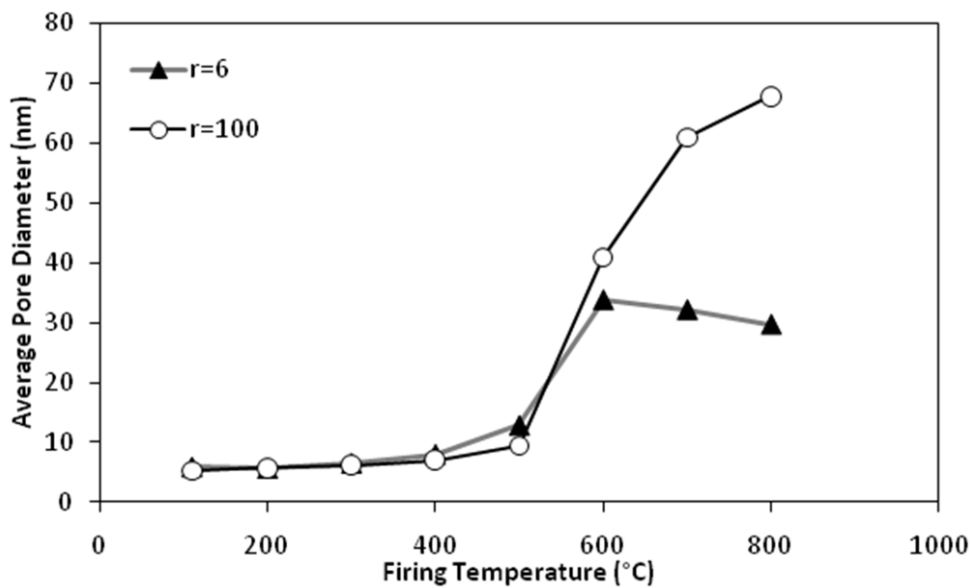

**Fig.14. Average pore diameters determined by BJH analysis**

## 4. Discussion

### 4.1. Organic content

In the present work, commonly employed hydrolysis catalysts were not used. Hence, the hydrolysis at r = 6 was incomplete at room temperature. The carbon retained in the precipitates was still present after drying at 110°C and the associated removal of residual volatile compounds. This is evident from the presence of the C-H vibrations in ATR-FTIR and laser Raman microspectroscopy and from the elemental microanalysis results. The absence of C-H vibrations in the ATR-FTIR and Raman spectra from powders fabricated through aqueous precipitation suggests that in these samples hydrolysis was complete, although carbon may have been present as an adsorbate on $TiO_2$ surfaces resulting from the combustion of ROH species during drying.

The requirement of a significant excess of water for complete hydrolysis of titanium alkoxides has been reported previously [17, 49]. However, it is likely that the hydrolysis level adequate for complete hydrolysis is dependent on the peptisation and pH levels. The presence of retained organic groups in the precipitates is likely to impact on the morphology, phase composition, and performance of $TiO_2$ powders prepared by the hydrolysis of alkoxides.

### 4.2. Precipitate Phase Formation

As shown by XRD analysis, hydrolysis at r = 100 caused crystalline precipitation of anatase and brookite, which is likely to be a result of a higher level of supersaturation of the solution by water and the consequent incipient formation of larger nuclei [25]. In contrast, r = 6 results in precipitation of an amorphous phase, most likely due to lower supersaturation and consequent incipient formation of nuclei smaller than one unit cell, the minimal nucleus size required for crystalline precipitation. It also is possible that the retention of organic groups contributes to prevention of the formation of the ordered anatase lattice. This phase formation behaviour is consistent with other reports of precipitate formation in titanium alkoxides [29, 50].

### 4.3. Morphology

From the analysis of the SEM and TEM images, r = 100 powders fired below 500°C exhibited larger grains in comparison with r = 6 powders. These grains were shown by TEM to be comprised of nano-scale crystallites which too were larger in r = 100 powders. This difference in crystallite size was less evident from XRD analysis as the coincidence of the anatase and brookite peaks in the r = 100 samples may have led to peak broadening and a lower resultant crystallite size determined by the Scherrer equation [44]. Since the growth of precipitates typically is controlled by processes at the particle/solution interface, the formation of larger grains in the r = 100 powders probably results from the lower free energy of crystalline solids in comparison to amorphous ones [48]. That is, the interface-controlled growth of crystalline precipitates would be expected to occur at more rapid growth kinetics compared with those in amorphous precipitates.

Powders prepared by precipitation in a predominantly aqueous medium exhibited a greater extent of agglomeration and lower extent of porosity. This is a result of a more rapid precipitation at r = 100 coupled with the denser particle packing that crystalline particles exhibit during drying in comparison with amorphous particles [51]. As shown in Fig. 14 and Fig. 1, pores in r = 100 powders were larger than those in r = 6 powders following the transformation to rutile. This is attributed to the aforementioned densification of agglomerates.

### 4.4. Phase transformation

The powders precipitated at r = 100 show a mixture of anatase and brookite. The ratio of anatase to brookite these powders has not been determined as the method of Spurr and Myers [40] is applicable only to anatase + rutile mixtures. Using the modified version of this method, substituting $X_A + X_B$ for $X_A$, the analysis shows that the transformation of anatase to rutile in r = 6 appeared to be slower than that of anatase + brookite to rutile in r = 100 powders. The presence of brookite in r=100 powders is likely to have enhanced the formation of rutile in these powders through a higher density of phase interfaces and consequently higher interfacial energy. This is supported by numerous publications in which brookite has been reported to act as a nucleation site for rutile formation in anatase [51-54]. This also may explain the phase transformation onset temperature of ≤500°C observed in r = 100 powders, a temperature lower than what is generally reported as the onset temperature of the anatase to rutile phase transformation [45].

The more rapid rutilation of r = 100 powders could also be the result of larger metastable crystallites transforming more readily to the equilibrium rutile phase due to enhanced growth kinetics. Conversely, the higher level of retained carbon in samples prepared by precipitation in organic media may have enhanced the anatase to rutile phase transformation in r = 6 through increased levels of oxygen vacancies as the carbon may act as a reducing agent. Increased levels of oxygen vacancies also would facilitate the structural rearrangement involved in the anatase to rutile phase transformation [45], and offset the slower transformation kinetics in the absence of brookite.

### 4.5. Surface area

The surface area of powders prepared in this work compare favourably to alkoxide-derived powders synthesised in other work [13, 18], as well as commercially available Degussa P-25 powder [1, 55], although powders prepared by hydrolysis of $TiCl_4$ generally show higher surface areas [56]. As shown in Fig. 12 and Fig. 13, a sharp reduction in surface area accompanied the anatase to rutile phase transformation at ~600°C. This effect has been reported elsewhere [57-60] and results from the previously described grain growth associated with the anatase to rutile and the brookite to rutile phase transformations. The reduction in surface area is further evident from the decreased levels of terminating hydroxyl groups observed in ATR-FTIR analysis, shown in Fig. 6. As shown in the insert in Fig. 13, the reduction in surface area accompanying the anatase + brookite to rutile phase transformation is more significant in powders precipitated at r = 100 and the surface area of rutile powders fabricated this way exhibit low surface areas around 0.5-2 $m^2 g^{-1}$. Such low surface areas are undesirable from the photocatalysis perspective.

At lower temperatures, the variation in surface areas is consistent with that of the grain sizes while, at higher temperatures, the decrease in surface area is larger in r = 100 powders despite a more moderate increase in grain size. Again, this resulted from the densification of agglomerates, as emphasised in Fig. 1c. The lower surface areas of samples precipitated at r = 100 and heated at 110°-400°C is a result of the aforementioned faster nucleation and growth of crystalline precipitates and denser packing of these particles. The denser grain packing in r = 100 powders also explains the increase in the average pore diameter shown in Fig. 14 as small pores become closed during densification.

As crystallinity is required for photocatalytic activity in $TiO_2$ [61, 62], the high surface area of amorphous powders is not likely to enhance photocatalytic performance in such powders. The highest photocatalytic activity would be expected in powders synthesised at r = 100 heated at 110°C. Under these conditions, a comparatively high surface area in conjunction with the presence of a mixed anatase-brookite phase composition are expected to yield enhanced photocatalytic performance.

## 5. Conclusions

- Owing to the dimensions of incipiently formed nuclei relative to the dimensions of the $TiO_2$ unit cell, the formation of crystalline phases by precipitation from titanium isopropoxide solutions requires a significant excess hydrolysis ratio with precipitation in aqueous medium resulting in fully hydrolysed crystalline precipitates while precipitation in a predominantly organic medium, with a $H_2O$:Ti ratio of 6:1, results in an amorphous, incompletely hydrolysed $TiO_2$ product.

- Mixed anatase/brookite/rutile $TiO_2$ powders, with potentially attractive photocatalytic properties, can be formed by thermal treatment at ~500°-600°C of powder precipitated with a significant excess of water.

- Confirming previous studies, the transformation to rutile is enhanced in the presence of a secondary brookite phase in anatase as a result of interfaces with this phase acting as nucleation centres for rutile formation.

- As a result of better dispersion of precipitates, surface areas are approximately 30-60% higher in amorphous powders obtained by precipitation at lower hydrolysis ratios and fired at temperatures up to 300°C, relative to the more agglomerated crystalline powder from precipitation in aqueous media. Levels of surface area decrease significantly with the formation of rutile at 600°-800°C.

- As amorphous materials is not reported to exhibit photocatalytic activity, the use of higher hydrolysis ratios may be advantageous for the fabrication of photocatalysts due the formation of high surface area crystalline precipitates at low temperatures.